\documentclass{vgtc}                          

\graphicspath{{figures/}{pictures/}{images/}{./}} 

\usepackage{times}                     

\usepackage{tabu}                      
\usepackage{booktabs}                  
\usepackage{lipsum}                    
\usepackage{mwe}

\usepackage{mathptmx}                  

\onlineid{0}

\vgtccategory{Research}
\usepackage{tabularray}
\usepackage{multirow}
\usepackage{adjustbox}
\vgtcinsertpkg

\title{Gauging the Competition: Understanding Social Comparison and Anxiety through Eye-tracking in Virtual Reality Group Interview}

\author{Shi-Ting NI\\ %
        \scriptsize Hong Kong University of Science and Technology(Guangzhou) %
\and Kairong FANG\\ %
     \scriptsize Hong Kong University of Science and Technology(Guangzhou) %
\and Yuyang WANG\thanks{e-mail:yuyangwang@hkust-gz.edu.cn}\\ %
     \scriptsize Hong Kong University of Science and Technology(Guangzhou) %
\and Pan HUI\\ %
      \scriptsize Hong Kong University of Science and Technology(Guangzhou) }
\teaser{
  \centering
  \includegraphics[width=\linewidth]{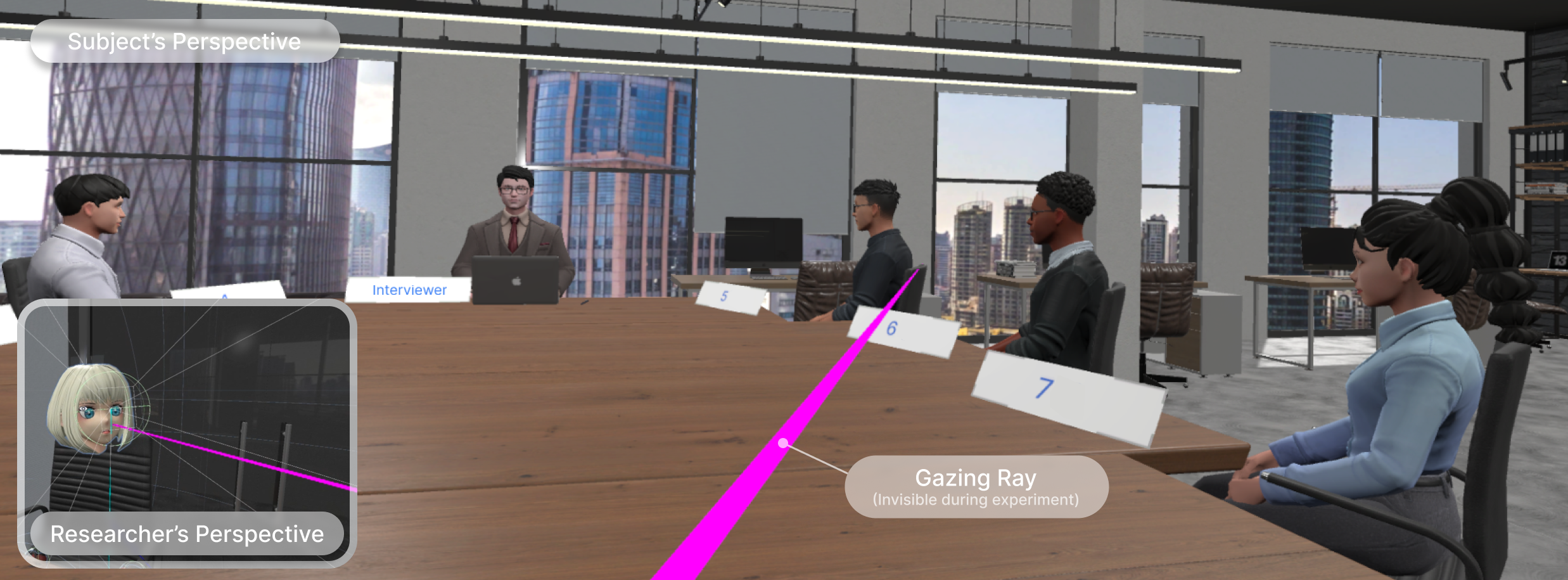}
  \caption{Immersive virtual reality interview scenario. The large window is the subjects' perspective; they can observe one interviewer and seven competitors in the experiment. The small window is the researcher's perspective, and the purple ray is the subject's sight line. Subjects could not see themselves (except their hands) or the ray during the experiment.}
  \label{fig:teaser}
}

\abstract{
    Virtual Reality (VR) is a promising tool for interview training, yet the psychological dynamics of group interviews, such as social comparison, remain underexplored. We investigate this phenomenon by developing an immersive VR group interview system and conducting an eye-tracking study with 73 participants. We manipulated peer performance using ambiguous behavioral cues (e.g., hand-raising) and objective information (public test scores) to measure their effect on participants' attention and self-concept. Our results demonstrate a "Big-Fish-Little-Pond Effect" in VR: an increase in high-achieving peer behaviors heightened participants' processing of social comparison information and significantly lowered their self-assessments. The introduction of objective scores further intensified these comparative behaviors. We also found that lower perceived realism of the VR environment correlated with higher anxiety. These findings offer key insights and design considerations for creating more effective and psychologically-aware virtual training environments that account for complex social dynamics.
} 


\begin{document}


\firstsection{Introduction}

\maketitle

Virtual reality (VR) offers researchers precise control over complex social scenarios while providing users with immersive, realistic experiences~\cite{pan2018and}. VR interview simulations, in particular, are a promising way to alleviate interview anxiety. For example, the Job Interview Simulation Training (JIST) application significantly reduced the heart rates of unemployed individuals during interviews, suggesting that familiarity with the interview process can alleviate anxiety~\cite{aysina2017using}.

VR interview training has enhanced performance across diverse populations, including individuals with autism spectrum disorder~\cite{adiani2022career}, veterans and former offenders~\cite{hartholt2019virtual}, young people not in education, employment, or training (NEET)~\cite{gebhard2014exploring}, and highly shy students~\cite{jin2019developing}. Moreover, completing more virtual interviews correlates with higher employment rates~\cite{smith2017mechanism}.

Beyond training, VR is also emerging as a talent selection tool. For instance, Lloyds Bank used VR for initial job interviews as early as 2016 to identify skilled digital and IT professionals~\cite{mixed2021virtual}. Similarly, technologies like Meta's Horizon Workrooms, with its gesture-based interaction, enable more natural and expressive VR interview scenarios~\cite{facebook2021horizon}.

While many studies confirm that VR exposure can improve interview skills and that objective factors like question type~\cite{luo2024using}, visuals~\cite{kwon2013level}, and interviewer attitude~\cite{kwon2009study} affect anxiety, few have investigated the underlying psychological mechanisms affecting a candidate's performance. Furthermore, existing research primarily focuses on one-on-one interviews, neglecting the common group interview format where multiple candidates are interviewed simultaneously~\cite{airswift2024group}. In these scenarios, candidates face pressure from both interviewers and peers. It remains unclear how a peer's strong performance might induce anxiety in a candidate and consequently affect their self-evaluation and performance. This research gap limits the use of VR for anxiety desensitization across different interview settings and hinders its potential as a comprehensive talent selection tool.

Research suggests a causal link between social anxiety and social comparison. Anxious individuals often adopt a competitive schema, leading to more upward comparisons, a perceived disadvantage, and negative self-evaluation~\cite{mitchell2014experimental}. Such comparisons can create dysfunctional self-beliefs that increase depression and anxiety~\cite{mccarthy2020exploring}. As a competitive social behavior, job interviews may also trigger social comparison, which is how individuals evaluate their abilities and opinions without objective standards~\cite{festinger1954theory}.

People typically compare themselves to similar peers to gauge their competence~\cite{suls2000selective}. This behavior is often categorized as upward or downward comparison and can be measured by a Social Comparison Orientation (SCO) score~\cite{gibbons1999individual}. In competitive settings, this can lead to the Big-Fish-Little-Pond Effect (BFLPE): individuals in high-achieving groups tend to have lower self-concepts than peers of similar ability in less capable groups~\cite{marsh1987big}. However, individuals with higher self-concepts often receive more favorable first impressions~\cite{riggio1986impression}. Thus, whether social comparison induces anxiety or leads to underestimation, it can negatively impact a candidate's mindset and performance.

Therefore, we developed a VR group interview system to observe candidates' behaviors and explore their psychological mechanisms. Previous work shows that gaze features and pupil diameter can objectively reveal social comparison mechanisms~\cite{halszka2017eye}. We created an immersive VR group interview scenario incorporating social comparison cues. Using eye-tracking, we collected participants' attentional data during the interview. We combined this data with self-reports to investigate how social comparison influences anxiety and self-evaluation.

We conducted a controlled experiment by manipulating the presence of objective comparison information (e.g., test scores). This design allows us to investigate two key aspects: first, whether individuals engage in social comparison without explicit benchmarks~\cite{hasenbein2023investigating}, and second, how objective benchmarks influence the extent and outcome of such comparisons~\cite{wood1996social}. We also collected data on participants' comparison tendencies to explore their relationship with interview anxiety and SCO.

Our goal is to understand the psychological mechanisms of candidates in VR group interviews. We aim for our findings to inform the design of VR interview systems, ultimately improving both user experience and recruitment effectiveness.

In summary, we propose the following research questions (RQs):
\begin{itemize}
\item[RQ1] Does a VR interview trigger job seekers to engage in social comparison actively?
\item[RQ2] How does social comparison influence job seekers’ self-assessment and interview anxiety?
\item[RQ3] Does objective comparison information enhance job seekers’ social comparison?
\item[RQ4] Is the level of VR interview anxiety among job seekers associated with their SCO levels and comparison preferences?
\end{itemize}

To investigate these questions, we developed a realistic VR group interview in Unity3D (Fig.~\ref{fig:teaser}) and used an HTC Vive Pro Eye HMD to collect eye-tracking data. We analyzed this physiological data alongside self-reported measures, combining experimental control with the complexity of real-world settings~\cite{hasenbein2023investigating,scorolli2023would}. Our approach aims to uncover the psychological mechanisms of individuals in VR interviews. By analyzing physiological and subjective feedback, we also discuss potential design improvements for VR interview systems to create a more user-centered experience for both training and talent selection.

\section{Related Work}\label{sec:related work}

\subsection{Consistency between VR and Real-World Social Interactions}

Numerous studies show that VR can effectively replicate real-world human behaviors in experimental research. The validity of using VR for social and productive activities is increasingly supported by empirical results that align with established theories. For social behavior studies, VR provides maximal experimental control over complex scenarios and enables testing infinite combinations of social variables~\cite{pan2018and}.

For example, one study on moral behavior used VR to present realistic virtual events, highlighting that immersive VR combines the controlled aspects of questionnaires with the realism of daily life~\cite{thorn2020investigation}. Other research has validated VR's reliability in replicating physical environments to study occupant comfort~\cite{alamirah2022immersive}. In one case, researchers compared patients' experiences in a VR model of a real building with their experiences in the actual building. They found strong similarities across thoughts, emotions, and sensations, suggesting VR can serve as a valid proxy for real-world experiences~\cite{riviere2024towards}.

Luo confirmed that VR-based interviews replicate real-world experiences, inducing similar or even greater anxiety levels~\cite{luo2024using}. This consistency extends beyond situational responses; people also react to avatars as they would to real people~\cite{garau2005responses}. For instance, a virtual audience can induce anxiety in individuals with public speaking fear just as effectively as a real one~\cite{robillard2010using}. This finding is critical for replicating social interactions in VR~\cite{pan2018and} and confirms that such interactions can yield performance equivalent to real-world scenarios.

\subsection{Feasibility of VR in Symptom Management}

VR's immersive nature creates significant opportunities for non-pharmacological interventions, making it an accessible and cost-effective healthcare solution~\cite{matsangidou2024transported}. Its feasibility has been demonstrated in managing dementia symptoms~\cite{matsangidou2024transported}, various types of pain~\cite{austin2022feasibility,ahmadpour2019virtual}, eating disorders~\cite{matsangidou2022now}, substance use~\cite{albright2021innovative}, Parkinson’s motor symptoms~\cite{goh2021video}, and social impairments in children with Autism Spectrum Disorder~\cite{didehbani2016virtual}.

VR is also an effective alternative for Exposure Therapy (ET), a common treatment for anxiety and phobias~\cite{carl2019virtual,matsangidou2022now}. It overcomes the visualization challenges of imaginal exposure and achieves results comparable to in-vivo exposure at a much lower cost~\cite{maples2017use}. In one experiment, students who trained in a VR clinic with virtual patients reported less exam anxiety and higher self-efficacy during subsequent clinical examinations~\cite{concannon2020immersive}. Similarly, VR interview systems have been shown to increase psychological preparedness and reduce anxiety for unemployed individuals~\cite{aysina2017using} and those with autism~\cite{adiani2022career}.

While VR interventions for interview anxiety are not widely studied, VR is considered well-suited for ET for anxiety disorders. This suitability stems from VR's ability to induce presence and immersion to evoke fear~\cite{fallon2021multi,luo2024using}, while also allowing for repeated, safe, and modifiable exposure that maintains real-world applicability~\cite{maples2017use,ahmadpour2024building}.

\subsection{Social Anxiety and Social Comparison}

Job interviews are common social situations~\cite{heimberg1995social} where social comparison frequently occurs~\cite{wheeler1992social}. Social comparison is a core human behavior where individuals identify similarities with a target or focus on differences to distinguish themselves~\cite{wood1989theory}. When pursuing a goal, people seek specific comparison targets to achieve a desired outcome~\cite{wood1996social}, leading to upward or downward comparisons. An individual's tendency for comparison can be measured with the Social Comparison Orientation (SCO) score~\cite{gibbons1999individual}. Those with high SCO scores compare themselves more often~\cite{buunk2005social}, are more influenced by these comparisons~\cite{gibbons2002drinking}, and tend to react more negatively to upward comparisons~\cite{lockwood2002could}.

Research shows that Social Anxiety Disorder (SAD) is linked to more frequent social comparison. Individuals with SAD often adopt a competitive social schema, perceive themselves as disadvantaged, and engage in negative self-evaluation—a process that may perpetuate the disorder~\cite{mitchell2014experimental}. Compared to the general population, SAD patients report more upward social comparisons, focusing on personality and social skills~\cite{antony2005social}. Mitchell found that men with severe social anxiety exhibited more negative self-evaluations when paired with high-performing peers versus average ones~\cite{mitchell2014experimental}, a phenomenon similar to the BFLPE~\cite{marsh1987big}.

The BFLPE has been validated across different grade levels, countries, and cultures~\cite{holm2020big,gilbert2022can,yuan2023conformity,marsh2003big}. A key indicator of the BFLPE is when an individual's self-evaluation is inversely proportional to their group's average performance, suggesting social comparison has negatively impacted their self-concept~\cite{fang2018big}. Therefore, an interviewee's anxiety and self-evaluation may stem from social comparison, potentially leading to inaccurate self-assessment due to the BFLPE and ultimately affecting their interview outcome.

Therefore, improving VR interview systems to better simulate real-world scenarios or serve as recruitment platforms holds significant value. By understanding the psychological mechanisms behind interview anxiety, we can enhance VR training efficacy and foster more equitable talent evaluation.

\section{Method} 
\label{sec:experiment setup}

\subsection{Experiment setup}

We developed an immersive virtual environment for the experiment, simulating a computer-related group interview scenario. The materials are from real job market questions for computer engineers or programmers. The experiment consists of three stages: a written test, and two rounds of interviews. Participants complete a 10-question test based on computer science, but they are given a predetermined score instead of their actual test results. The two rounds of VR interviews include mandatory and quick-response questions of equivalent difficulty. Participants are asked to introduce themselves to virtual competitors~\footnote{NPC controlled by LLM} to enhance presence and engagement. In the quick-response session, we use the \textit{number of people raising hands} as an ambiguous comparison indicator. Only the quickest two hand-raisers, including participants and virtual NPC, can answer questions in each round. The selected virtual NPC provides nearly perfect answers, suggesting that raising hands indicates confidence in answering correctly.

\begin{figure}
    \centering
    \includegraphics[width = 1\linewidth]{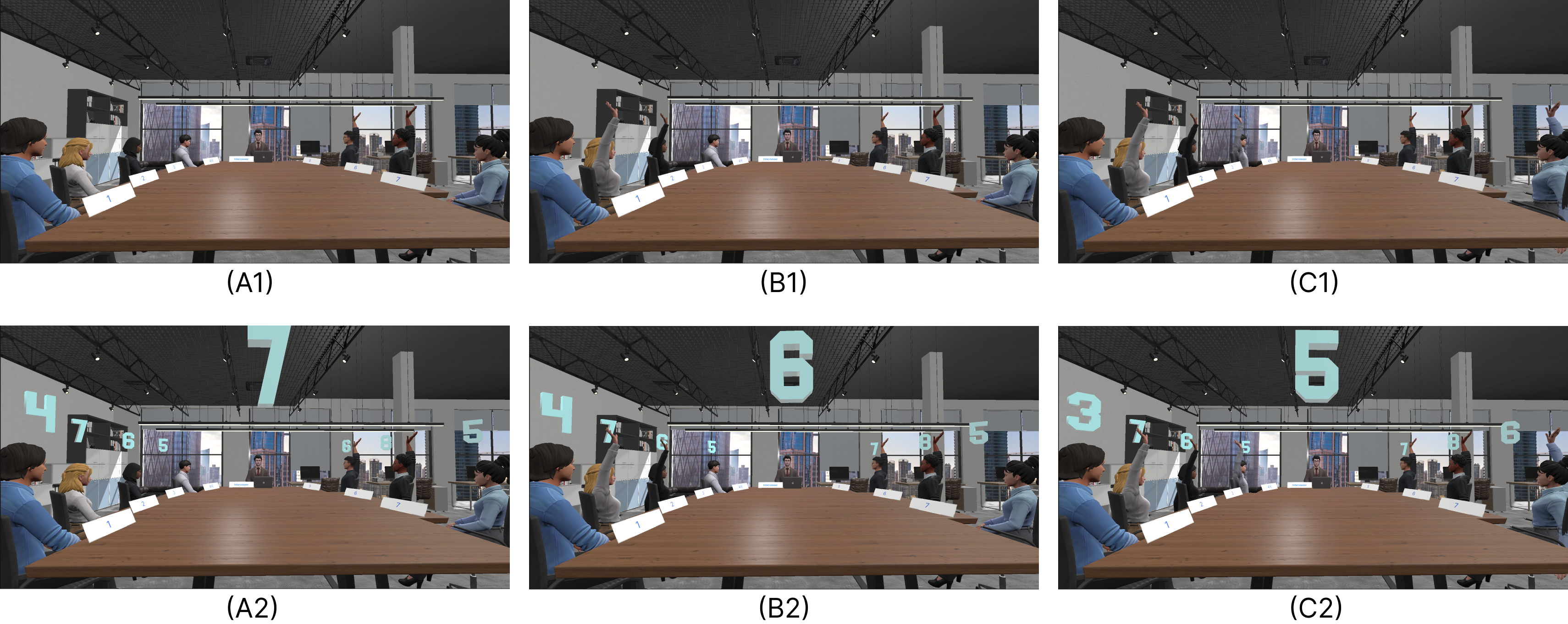}
    \caption{Immersive virtual reality scenes corresponding to the Round 1 (top) and Round 2 (bottom) quick-response sessions under the three experimental conditions.}
    \label{fig:scenes}
\end{figure}

The experiment has three conditions based on the number of people raising their hands: two, four, and six. Participants are randomly assigned to these conditions without prior notice. The second round of interviews differs from the first by the addition of \textit{public written test scores} as an objective comparison indicator. In this round, participants can see the written test scores of all interviewees, including their own, displayed above their respective avatars. These scores, however, are not actual test scores but assigned ranks as psychological hints to the participant, with scores of 7, 6, and 5 points under the conditions of 2, 4, and 6 hand-raisers, respectively.(Fig.~\ref{fig:scenes}).

\subsection{Experiment condition}
This study includes two independent variables: \textit{number of people raising hands} and \textit{whether written test scores are public}. Hand-raising, a clear yet somewhat ambiguous ability indicator~\cite{hasenbein2023investigating}, is adopted for its appropriateness in interview scenarios, particularly in quick-response segments. According to our design, virtual competitors raising their hands and answering questions perfectly lead participants to equate hand-raising with high ability. We controlled competitors' activity level through three conditions: 2, 4, and 6 people raising their hands. Two people raising their hands imply low competitor activity, positioning participants as the "big fish" in the group. Conversely, with six people raising their hands, participants are positioned as the "small fish." The first interview round serves as a control for the second variable. In the second round, each interviewee's test score is publicly displayed as an objective comparison indicator. Participants' scores are above average when they are the "big fish." Displaying scores instead of rankings explores participants' social comparison initiative. 

The study's dependent variables are participants' eye movement data and self-evaluation. In tracing Bell's work on eye movement and the nervous system, eye-tracking technology can record eye movement changes and gaze positions~\cite{bell1823xv}. This technology allows for measuring human interaction processes under challenging conditions~\cite{valtakari2021eye}, tracking gaze direction and sequence within a specific timeframe. The eye-mind link~\cite{rayner2015evidence} makes eye-tracking a reliable tool for studying attention allocation~\cite{carter2020best}, highlighting the potential of gaze characteristics and pupil diameter in understanding social comparison mechanisms~\cite{halszka2017eye}. Research suggests that gazing at peers indicates that students are paying attention to comparative information in the scene~\cite{carrasco2011visual}, gaze frequency and duration suggest active processing of social comparison information~\cite{mason2013eye}. At the same time, pupil diameter often reflects the level of concern for the information~\cite{maier2021pupil}. To explore the participants' attention to and concern for comparative information, we selected the \textit{gaze number}, \textit{gaze frequency}, \textit{gaze duration}, and \textit{pupil diameter} during the interview as metrics.

Given the complexity of social comparison mechanisms, we included participants' gender, personal achievements, and social orientation in the self-report, as these factors are considered significant in this field~\cite{hasenbein2023investigating}. Personal achievements are represented by participants' past academic performance (latest math, English, and overall scores) and self-assessed intelligence. We gathered self-assessments of participants' interests and proficiency in computer science, along with their actual test scores, to evaluate achievements related to the interview content. Additionally, we recorded participants' anxiety levels after the VR interview and the reference individuals they used for social comparison during the interview.

\subsection{Platform development}

\begin{figure}
    \centering
    \includegraphics[width = 1\linewidth]{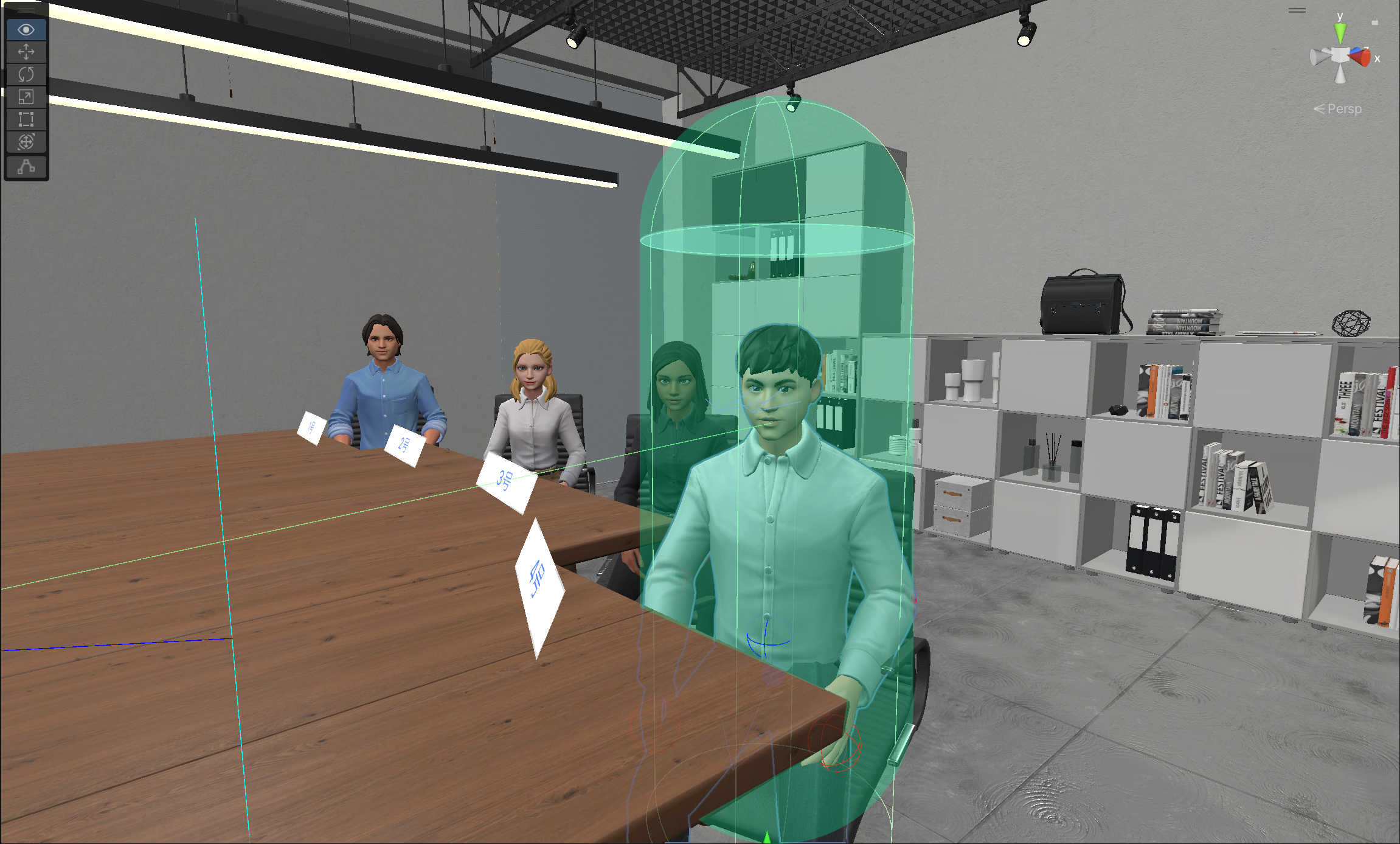}
    \caption{Capturing participants' gaze targets by setting up colliders (based on the researcher's perspective).}
    \label{fig:collider}
\end{figure}

The immersive VR interview scenario used in this study was developed with the Unity game engine and streamed to an HTC VIVE Pro Eye VR headset via SteamVR. Eye movement and pupil diameter data were collected using the integrated Tobii eye tracker in the headset. The participants’ VR view was projected in real time onto a local computer monitor. However, participants were unaware that their eye movement data was being collected during the interview and could not see their gaze orientation visualized as a purple ray (Fig.~\ref{fig:teaser}). To accurately identify the specific competitor participants were gazing at during a certain period, we set up a collider around each virtual competitor's body. If the participant's gaze ray intersects the collider for more than 500ms \cite{hasenbein2023investigating}, it was recorded as gazing at the competitor once (Fig.~\ref{fig:collider}). Metrics such as the number, frequency, and duration of gazes at virtual competitors, as well as pupil position and diameter, were output in tabular form. To avoid disrupting participants' normal interview behavior due to observing their virtual hand movements', their virtual hands were rendered invisible.

The interview scene was set in a realistic office environment featuring seven virtual competitors (interviewees) and one interviewer. Human-shaped avatars were chosen for the competitors because they provide a stronger sense of presence than non-human avatars (e.g., animals or geometric shapes)~\cite{girondini2023speaking}. While differences in anxiety levels between cartoonish human avatars and realistic human avatars are minimal~\cite{kwon2013level}, overly realistic avatars may trigger the uncanny valley effect, potentially interfering with measurements due to discomfort~\cite{seyama2007uncanny}. Therefore, the avatars were designed with human-like characteristics using Ready Player Me, with variations in skin tones, facial features, genders, and appropriate interview attire.

To make the avatars’ behavior more realistic, Sony Mocopi was used for motion capture, and limb animations were refined with animation masks and Adobe Mixamo. Eye movements, such as blinking at random intervals and gazing at the interviewer, were controlled through custom C\# scripts. Since mismatched or unrealistic voices can reduce social presence and elicit negative emotional responses~\cite{higgins2022sympathy}, the voices of all virtual competitors were synthesized using advanced text-to-speech technology. OVR lipSync was employed to animate lip movements in real time, and the pitch, tone, speed, and emotion of each voice were tailored to match the avatar’s gender, appearance, and dialogue.

To enhance the realism of the interview content, the speech of the virtual competitors and the interviewer was carefully scripted in conversational language after being reviewed for accuracy by computer experts. Each virtual character was assigned a distinct personality and speaking style to create a vivid and authentic interview experience.

\section{User Study} \label{sec:user study}

The experiment was designed to investigate whether participants engage in active social comparison during VR interviews, leading to increased anxiety and affecting self-assessment as described by the BFLPE. Additionally, we aimed to explore the influence of objective information on the degree of social comparison and the correlation between anxiety levels and social comparison tendencies. The hypotheses are as follows:

\begin{itemize}
    \item[H1] VR interviews trigger participants to actively engage in social comparison, with the degree of comparison increasing as the number of hand-raising virtual peers rises.
    \item[H2] Increased social comparison heightens interview anxiety and lowers self-assessment.
    \item[H3] The presence of objective comparative information in the VR interview environment enhances social comparison among participants.
    \item[H4] Participants with higher SCO and a preference for upward comparison experience greater anxiety in VR interviews.
\end{itemize}

\subsection{Participants}

We recruited university students through posters and announcements. Given the specialized nature of the interview content, we eventually selected 79 students with a background in computer science or computer art to participate in the experiment. A prior power analysis, assuming a medium effect size (f = .38) and 80\% power, indicated a need for at least 24 valid samples per group in three hand-raising conditions. Seventy-nine participants with expertise in computer science were recruited, but due to technical and data collection issues, six were excluded, leaving 73 participants (62 postgraduates, 41 males), each being compensated with a gift.  

\subsection{Measurement}
The self-report questionnaires used in this study incorporate scales for self-evaluation, intellectual self-evaluation, social comparison orientation, anxiety level, sense of realism, and presence. These scales, sourced wholly or partly from validated professional scales in social comparison research, are expertly translated into Chinese and adapted to the experimental context, maintaining the original evaluation levels.

\begin{itemize}
    \item Four items for self-evaluation are selected from the Self-Description Questionnaire (SDQIII)\cite{mukli2021albanian}, a validated tool developed for late adolescents based on Marsh's SDQ\cite{marsh1984self}. The SDQIII contains 13 scales with 136 items, about half of which are negatively described.
    \item Four items for intellectual self-evaluation are selected from Andrea's Intellektuelle scale, a tool developed according to SDQIII for measuring young people's self-concept. This scale has been validated in multiple university student samples~\cite{schwanzer2005entwicklung}.
    \item Eleven items from the Iowa-Netherlands Comparison Orientation Measure (INCOM) are used to assess social comparison orientation, which includes two measurement dimensions: ability comparison and opinion comparison. The INCOM scale has been validated in large samples in the Netherlands and the United States and is considered a reasonable and effective tool for measuring social comparison orientation~\cite{gibbons1999individual}.
    \item Six items from the Measure of Anxiety in Selection Interviews (MASI) to measure participants' anxiety levels during the IVR interview. The MASI is a commonly used tool for assessing interview anxiety~\cite{mccarthy2004measuring}.
    \item Eight items from the Presence Questionnaire (PQ) are introduced to measure participants' sense of realism and presence in the virtual environment~\cite{witmer1998measuring}.
\end{itemize}


\subsection{Experiment procedure}

This study has been approved by the Ethics Committee of the university. Before the experiment, participants were informed of potential risks and provided written informed consent. The experiment was conducted in a transparent, soundproof room to avoid extra interruptions. Each one participated separately. The experimental procedure was designed as follows:


After being informed of the experimental process, participants filled out the first part of a self-report, including demographic information (such as \textit{gender, age, etc}.), self-assessment of academic ability (including \textit{math scores, language scores, and overall scores}), self-assessment of interest in computer science (such as "\textit{I am interested in computer science}"), self-assessment of computer science proficiency (such as "\textit{I understand professional knowledge in the field of computer science}" ), familiarity and adaptability to VR (such as "\textit{I am very familiar with using VR}" ), self-assessment of intelligence (such as "\textit{I often feel less intelligent than others}" ), and social orientation (such as "\textit{I don't often compare myself with others}" ), totaling seven scales. 

Participants completed a 10-minute, 10-question test on computer science, with scores recorded but undisclosed. 

The researchers then offered participants the choice to review experience posts related to computer science interviews (either passed or failed, claimed to be guides written by participants who had previously taken this interview). Participants could view none, one, or both, indicating their comparison inclination. Choosing to view the successful experience post indicated a willingness for upward comparison. Their choices were recorded and cross-referenced with subsequent scales and interviews.

\begin{figure}
    \centering
    \includegraphics[width = 1\linewidth]{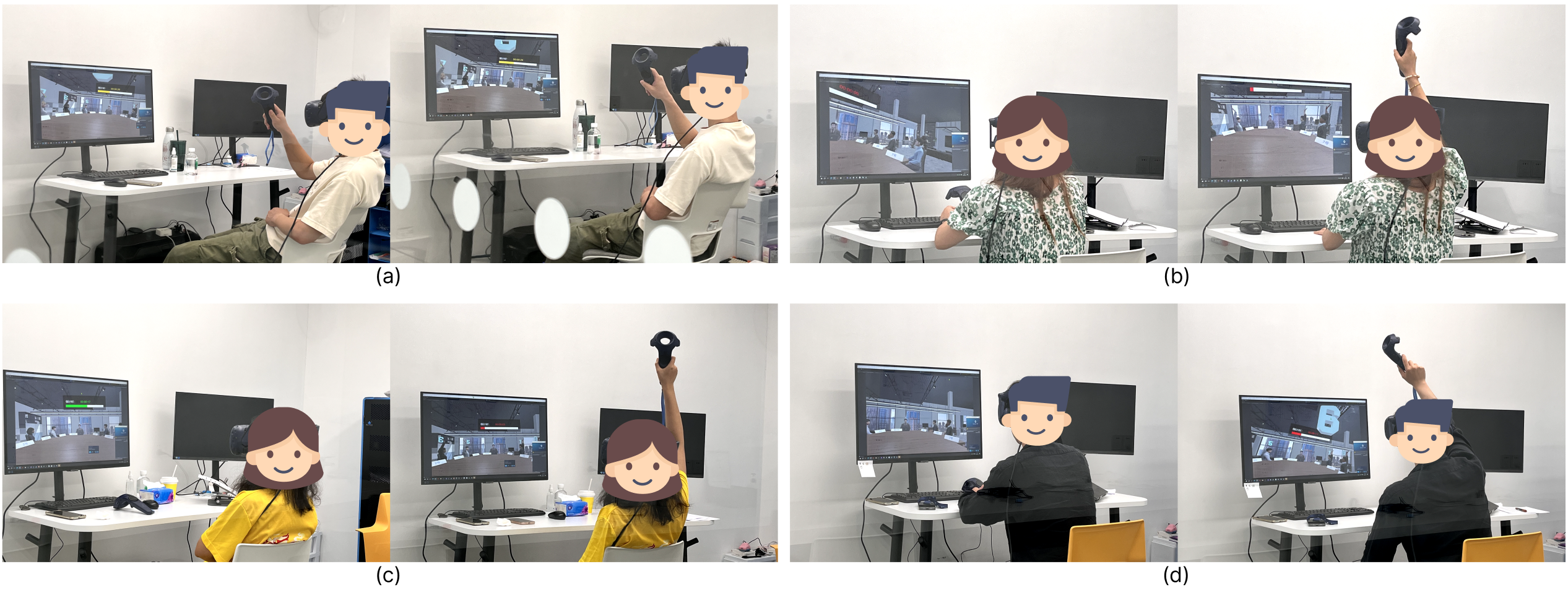}
    \caption{Real shots of the interview process, with the left side showing the resting state and the right side showing the hand-raising quick-response state.}
    \label{fig:Frame200}
\end{figure}

Participants were then informed of claimed test scores, assisted with donning the HTC Vive Pro Eye, and completed eye tracker calibration. They proceeded to a 15-minute, two-round interview. To ensure comfort and immersion, participants were informed of the room's soundproofing, with staff observing from outside and only entering during the break to assist with the second part of the self-report (including only one self-assessment scale with four questions, such as "\textit{Compared to other candidates, I am good at answering questions correctly}") and explain the meaning of numbers above avatars' heads in the second round (Fig.~\ref{fig:Frame200}).

After the VR part, participants were asked to fill out the third part of the self-report. This section assessed their anxiety levels during the IVR interview (e.g., "\textit{During the recent interview, my hands trembled involuntarily}"), self-evaluation (e.g., "\textit{Compared to other candidates, I am adept at answering questions correctly}"), upward comparison willingness (e.g., "\textit{I believe reading successful interview experiences before the interview will boost my motivation}"), sense of presence (e.g., "\textit{I felt that the interaction with the interview environment was very natural}"), sense of realism (e.g., "\textit{My experience in this virtual environment simulated that in the real world}"), and VR sickness (e.g., "\textit{I felt discomfort (dizziness, nausea, etc.) due to the equipment}"), totaling six scales. 

\begin{figure}
    \centering
    \includegraphics[width = 1\linewidth]{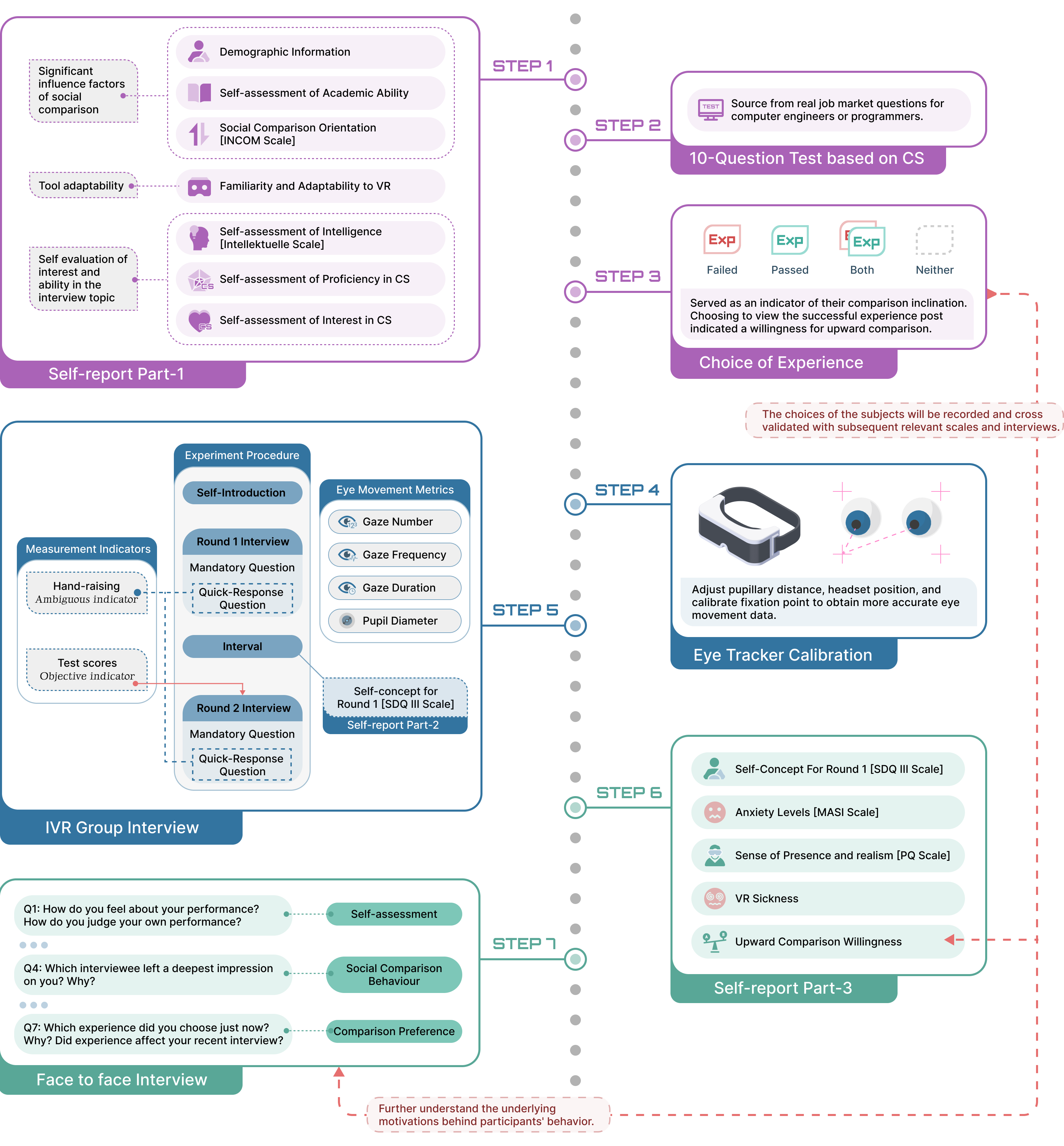}
    \caption{Flowchart of user study; the three colors successively represent the before, during, and after stages of the experiment, totaling 7 steps.}
    \label{fig:flowchart2}
\end{figure}

Following this, researchers conducted a follow-up interview lasting at least 10 minutes to delve deeper into the participants' behavioral motivations based on their recorded interview performance (Fig.~\ref{fig:flowchart2}).

\section{Results} \label{sec:results}

\subsection{Data pre-processing}

Eye movement data are collected via a Tobii eye tracker inside the VR headset, recording gaze number, frequency, duration, and pupil diameter during the experiment. We address data loss (mainly pupil diameter data) from equipment malfunctions using linear interpolation and smooth the data with a Savitzky-Golay filter (window length = 5, polyorder = 2)~\cite{savitzky1964smoothing}, normalizing it using divisive baseline correction with the first row of pupil data values collected as the baseline. Data are analyzed separately for each interview round, with log transformation applied to the \textit{gaze frequency}, \textit{gaze duration}, and \textit{pupil diameter} datasets to facilitate analysis.

We retain the maximum similarity to the original scales during measurement. However, in data processing, all negatively described items, except those in the intellectual self-evaluation scale, are reverse-coded to ensure total scores from the same scale reflect overall participant trends. Thus, higher scores affirm the scale's positive conclusions. However, the intellectual scale contains four negatively described items, so higher total scores on this scale indicate lower self-evaluation of intelligence.


\subsection{Descriptive statistics}

Analyzing 73 samples across three experimental conditions with descriptive statistics over the entire experiment duration, we sought to investigate how the activity levels of virtual competitors (2 vs. 4 vs. 6 competitors raising hands) influence participants' attention to and concern with comparative information, measured via eye movement metrics (detailed in Table~\ref{table1}). The results revealed a trend in gaze duration: participants in the two-competitor condition exhibited the longest gaze duration (M = 242.98, SD = 65.52), followed by those in the four-competitor (M = 232.22, SD = 68.35) and six-competitor conditions (M = 225.06, SD = 70.31). Additionally, the two-competitor condition showed significantly higher gaze frequency (M = 156.04, SD = 29.63) and pupil diameter (M = 1.07, SD = .14). Participants in the four-competitor condition reported the highest SCO scores (M = 33.33, SD = 6.49), the highest interview anxiety (M = 15.12, SD = 6.01), and the lowest self-evaluation (M = 13.27, SD = 4.65). 

We analyzed how dependent variables changed in response to the independent variable across two interview rounds under three experimental conditions. Introducing a clear objective indicator (publicly disclosed test scores) resulted in an overall increase in gaze metrics, particularly gaze frequency and gaze duration. This suggests that the objective indicator stimulated more active social comparison behavior among participants.

When personal and competitors' scores were revealed, participants in the two-hand-raising condition showed a significant increase in self-assessment, from a mean of 13.75 (SD = 3.80) to 14.92 (SD = 5.44). In contrast, participants in the four-hand-raising condition exhibited a notable decrease in self-assessment, dropping from a mean of 14.33 (SD = 5.38) to 12.21 (SD = 4.49). Meanwhile, participants in the six-hand-raising condition showed almost no change in self-assessment across the two rounds.

\begin{table*}[ht]
\centering
\caption{Descriptive statistics for dependent variables in different experimental conditions}
\label{table1}
\begin{adjustbox}{width=\textwidth,center} 
\begin{tabular}{cccccccc}
\toprule 
\multicolumn{2}{c}{\multirow{2}{*}{\textbf{Experimental condition}}} & \multirow{2}{*}{\textbf{N}} & \textbf{Gaze number} & \textbf{Gaze frequency} & \textbf{Gaze duration} & \textbf{Pupil diameter} & \textbf{Self-concept} \\ 
\cmidrule(lr){4-8} 
& & & \multicolumn{5}{c}{\textbf{M(SD)}} \\ 
\midrule 
\multirow{3}{*}{\begin{tabular}[c]{@{}c@{}}\textbf{Without test scores} \\ \textbf{(1st round interview)}\end{tabular}} & \textbf{2 (25\%)} & 24 & 6.38(.71) & 54.79(13.88) & 83.87(27.64) & 1.07(.14) & 13.75(3.80) \\ 
\cmidrule(lr){2-8} 
& \textbf{4 (50\%)} & 24 & 6.79(.51) & 47.33(10.46) & 83.17(33.01)& 1.02(.13) & 14.33(5.38) \\ 
\cmidrule(lr){2-8} 
& \textbf{6 (75\%)} & 25 & 6.64(.49) & 54.40(16.73) & 81.04(33.25) & 1.01(.09) & 14.16(5.62) \\ 
\midrule
\multirow{3}{*}{\begin{tabular}[c]{@{}c@{}}\textbf{With test scores} \\ \textbf{(2nd round interview)}\end{tabular}} & \textbf{2 (25\%)} & 24 & 6.71(.62) & 71.08(17.60) & 110.60(41.20) & 1.08(.16) & 14.92(5.44) \\ 
\cmidrule(lr){2-8} 
& \textbf{4 (50\%)} & 24 & 6.67(.70) & 57.38(14.56) & 100.21(40.15) & 1.00(.13) & 12.21(4.49) \\ 
\cmidrule(lr){2-8} 
& \textbf{6 (75\%)} & 25 & 6.80(.41) & 67.52(18.84) & 98.40(34.20) & 1.00(.08) & 14.16(4.90) \\ 
\bottomrule
\end{tabular}
\end{adjustbox}
    \\
    \smallskip 
    \raggedright 
    {\small \textit{Note: M represents the mean, SD represents the standard deviation.}}
\end{table*}

\subsection{Analysis of variance (ANOVA)}

\begin{figure}[ht]
    \centering
    \includegraphics[width = 1\linewidth]{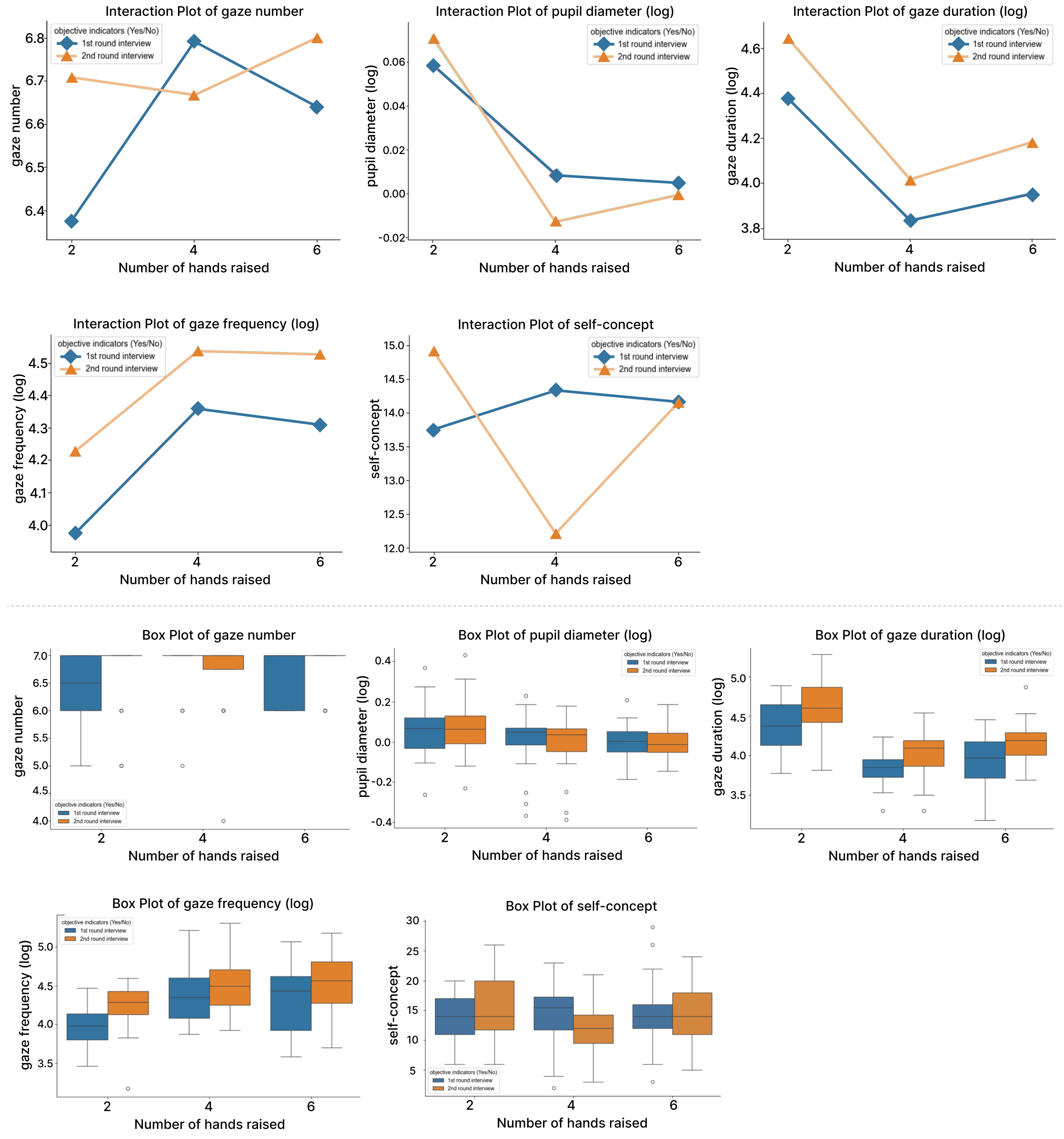}
    \caption{Interaction diagram analysis results (top) and box diagram (bottom) of Anova. Both of the independent variables have significant influences on gaze duration (log) and gaze frequency (log), and the cross terms of the independent variables have significant influences on pupil diameter (log) and self-concept. Independent variables had no significant effect on gaze number.}
    \label{fig:anova}
\end{figure}

This study included two independent variables and five dependent variables. To examine the effects of the independent variables \textit{number of people raising hands} and \textit{whether written test scores are public}, as well as their interaction on the dependent variables, we performed an ANOVA. Gaze frequency, gaze duration, and pupil diameter data were log-transformed due to small values and large variability. Since the variables were not normally distributed, an ART ANOVA was applied (Fig.~\ref{fig:anova}). The results (Table~\ref{table:Anova}) indicate that both independent variables have significant effects on gaze duration (log) ($F_{(2,70)}$ = 35.27, $F_{(1,70)}$ = 33.15, p $<$ .001) and gaze frequency (log) ($F_{(2,70)}$ = 9.85, $F_{(1,70)}$ = 28.24, p $<$ .001). In contrast, their interaction effects are not significant. For pupil diameter (log) and self-concept, changes due to single independent variables were not significant. However, their interaction showed significant effects, with F-values of 5.58 (p = .01) for pupil diameter and 3.68 (p = .03) for self-concept. Meanwhile, no significant effects were observed for gaze number across the independent variables or their interaction.



\begin{table}[h!]
\centering
\caption{ART ANOVA results for the independent and dependent variables}
\begin{tblr}{
  column{3} = {c},
  column{4} = {c},
  column{5} = {c},
  cell{1}{2} = {c},
  hline{1-2,17} = {-}{0.08em},
  hline{5,8,11,14} = {-}{0.05em},
}
\textbf{Dependent Variables}                    & \textbf{Source} & \textbf{F} & \textbf{p} & \textbf{np2} \\
                                                & Variable 1      & 1.26       & 0.29       & .04          \\
\textbf{Gaze~\textbf{\textbf{\textbf{number}}}} & Variable 2      & 2.2        & 0.14       & .03          \\
                                                & Interaction     & 2.55       & 0.09       & .07          \\
                                                & Variable 1      & 2.26       & 0.11       & .06          \\
\textbf{Pupil diameter (log)}                   & Variable 2      & 1.38       & 0.24       & .02          \\
                                                & Interaction     & 5.58       & .01**      & .13          \\
                                                & Variable 1      & 35.27      & .00***     & .50          \\
\textbf{Gaze duration (log)}                    & Variable 2      & 33.15      & .00***     & .32          \\
                                                & Interaction     & 0.38       & 0.69       & .01          \\
                                                & Variable 1      & 9.85       & .00***     & .22          \\
\textbf{Gaze frequency (log)}                   & Variable 2      & 28.24      & .00***     & .29          \\
                                                & Interaction     & 0.26       & 0.77       & .01          \\
                                                & Variable 1      & 0.39       & 0.68       & .01          \\
\textbf{Self-concept}                           & Variable 2      & 0.4        & 0.53       & .01          \\
                                                & Interaction     & 3.68       & .03*       & .10          
\end{tblr}
    \\
    \smallskip 
    \raggedright
    {\small \textit{Note: Variable 1 = Independent variable number of people raising hands, Variable 2 = whether written test scores are public, * means p$<$.05, ** means p$<$.01, *** means p$<$.001.}} 
    \label{table:Anova}
\end{table}

A post hoc analysis was conducted to examine inter-group differences for significant variables. Tukey’s HSD test was applied to normally distributed data, while Dunn’s test was used for non-normal data. No significant inter-group differences were found for the independent variable \textit{whether written test scores are public} concerning the dependent variables gaze number (p = .11), pupil diameter (log) (p = .62), and self-concept (p = .70). However, significant differences were observed for gaze duration (log) and gaze frequency (log), with p-values below .001. For the independent variable \textit{number of people raising hands}, no significant inter-group differences were found for gaze number and self-concept. However, a significant difference in pupil diameter (log) was observed between the two-hand-raising and six-hand-raising conditions (p = .009). Additionally, significant differences for gaze duration (log) and gaze frequency (log) were found between the 2 vs. 4 and 2 vs. 6 hand-raising conditions (p $<$ .001). No significant differences were observed between the 4 vs. 6 hand-raising conditions for gaze duration (p = .08) or gaze frequency (p = .91).

\subsection{Regression analysis}

Considering the post hoc analysis results, we used the data from participants in the condition where two competitors raised their hands as the reference group. Two crossed variables—\textit{4 vs. 2 competitors raising hands} and \textit{6 vs. 2 competitors raising hands}—were defined as new independent variables using dummy coding. The other independent variable was \textit{whether written test scores are public}. Considering individual differences, we included covariates such as gender, interview motivation, personal achievements, SCO levels, anxiety levels, upward comparison tendencies, actual and claimed written test scores, sense of realism, and sense of presence. Multiple regression analyses were conducted on the dependent variables to examine potential causal relationships between the independent and dependent variables.

For participants’ gaze frequency (log) toward virtual peers, results from the first interview round showed that an increase in the number of competitors raising hands significantly enhanced gaze frequency (p = .004 and p $<$ .001 for four and six competitors raising hands, respectively). The positive effect of six competitors raising hands ($\beta$ = .88) was more pronounced than that of four competitors raising hands ($\beta$ = .29). Additionally, claimed written test scores were significantly positively correlated with gaze frequency (p $<$ .001). In the second interview round, where objective comparison information (publicly disclosed claimed written test scores) was present, the significant positive effects of both raised hand conditions and claimed written test scores persisted (both p $<$ .001). Furthermore, participants with job-seeking or academic-related interview motivation also demonstrated significantly increased gaze frequency (p = .049).

For participants’ gaze number toward virtual peers, significant positive effects were observed across both interview rounds for the variables “4 competitors raising hands vs. 2 competitors raising hands,” “6 competitors raising hands vs. 2 competitors raising hands,” and “claimed written test scores” (all p $<$ .001). However, objective comparison information somewhat attenuated these effects, with $\beta$ values decreasing from 1.22 to .85 and from 2.11 to 1.86, respectively.

For participants’ gaze duration (log) toward virtual peers, the three variables mentioned above also showed significant positive effects in both interview rounds (all p $<$ .001). Additionally, objective comparison information slightly amplified these effects, with $\beta$ values increasing from .54 to .58 and from .96 to 1.21, respectively. In the first interview round, participants’ interview motivation had a significant negative effect on gaze duration (p = .02), indicating that participants with low interview-related motivation spent more time gazing at virtual competitors. Post-interview feedback suggested that these participants tended to adopt an observer role during the interviews, which may explain this behavior. In the second round, the introduction of objective comparison information eliminated the significance of this negative effect. However, gender differences emerged, with female participants exhibiting longer gaze durations toward virtual peers (p = .003).

For participants’ pupil diameter (log), the only factor with a significant effect was total academic achievement, which served as an indicator of personal accomplishments. Across both interview rounds, total academic achievement demonstrated a slight but significant negative correlation with pupil diameter (p = .013 and p = .008, respectively), regardless of objective comparison information.

The results indicate that the increase in virtual peers performing achievement-related behaviors (e.g., raising hands) significantly enhances students’ attention to comparison information (gaze number) and their deeper processing of it (gaze frequency and gaze duration). However, it does not have a significant effect on promoting their concern for comparison information (pupil diameter).

To further explore whether and how students’ processing of social comparison information affects their self-evaluation, we included the four dependent variables above as independent variables in multiple regression models, using self-concept as the dependent variable. The results show that in the first-round interview environment, which lacked objective comparison information, gaze duration (log) (p = .017) and pupil diameter (log) (p = .046) had significant negative effects on self-concept. This suggests that the longer participants spent processing comparison information and the more they cared about it, the more likely they were to lower their self-evaluation of their abilities. When objective comparison information was introduced in the second-round interview environment, gaze number had a significant positive effect on self-concept (p = .028), meaning that the more virtual peers students noticed during the second round of interviews, the more likely they were to give themselves a higher evaluation. Across both rounds of interviews, gaze frequency (log) showed no significant effect on self-concept. Among the covariates, participants’ self-assessment of their professional skills in computer science consistently exhibited significant positive effects on self-concept across all four regression models, suggesting that students’ self-evaluation in interviews is strongly grounded in their professional confidence within the relevant domain.

Notably, considering the overall data from the interview process, participants’ anxiety levels exhibited a negative correlation with their self-evaluation (measured by the average of self-concept scores after both interview rounds), but this correlation was not significant in any of the four models. To further investigate whether and how students’ processing of social comparison information affects their interview anxiety, we conducted analyses with participants’ post-interview reported anxiety levels as the dependent variable. Four regression models were set up, with gaze frequency (log), gaze duration (log), gaze number, and pupil diameter (log) as independent variables, respectively, using data collected throughout the entire experiment.
The results show that while most eye-tracking indicators had positive effects on anxiety levels, none of these effects were significant. However, intellectual self-assessment was a consistent and significant predictor of anxiety across all four models, with p-values of .031, .035, .034, and .026. Higher scores on intellectual self-assessment indicated that participants rated their intelligence lower, meaning that students who perceived themselves as less intelligent were more likely to experience anxiety during interviews. Additionally, participants’ realism scores for the IVR environment showed a consistent and significant negative effect, with p-values of .044, .04, .041, and .03. This suggests that participants who perceived the interview environment as less realistic were more likely to experience interview anxiety.

\begin{figure}[ht]
    \centering
    \includegraphics[width = 1\linewidth]{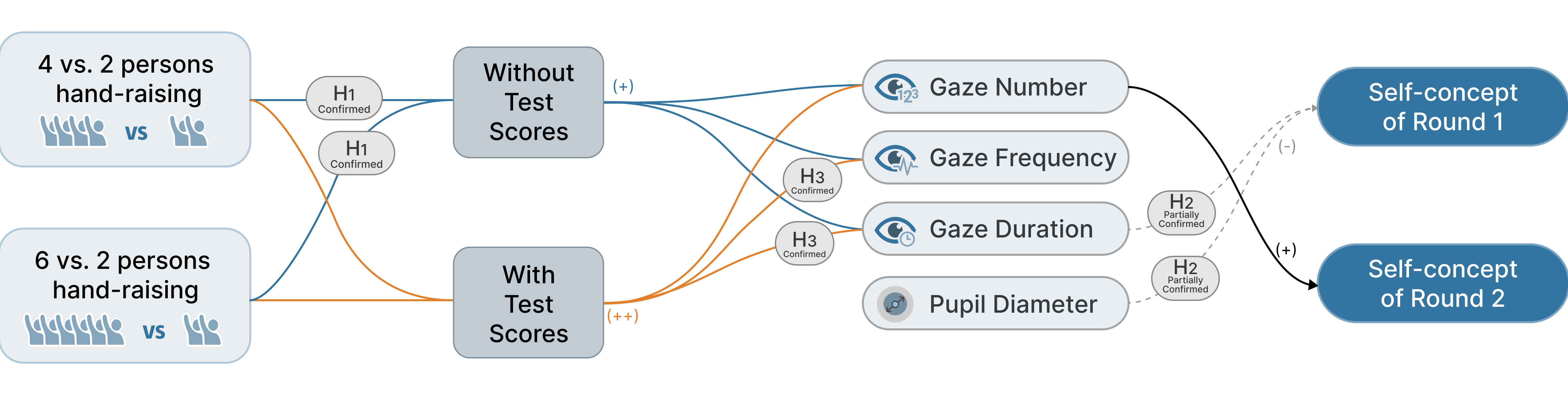}
    \caption{Results of regression analysis. Increasing the number of hand-raising promotes social comparison behavior and thus inhibits self-evaluation, while objective indicators heighten comparative behaviors and influence self-concept.}
    \label{fig:result}
\end{figure}

\section{Discussion}

The importance of interview training for job seekers is widely recognized, a belief rooted in common sense and supported by research~\cite{smith2017mechanism}. VR has shown significant potential for simulating and even directly conducting interviews~\cite{aysina2017using,smith2017mechanism,jin2019developing,mixed2021virtual,adiani2022career}. However, the current design of VR interview systems largely overlooks subjective factors that affect candidates in group interviews, particularly psychological processes. This oversight limits VR's ability to address the internal needs of individuals with limited interview experience.

To investigate the psychological processes of job seekers in VR-based group interviews, we developed an immersive virtual reality group interview scenario as a standardized yet realistic research environment. By combining eye-tracking data with self-reported measures, we examined spontaneous social comparison and its effects on anxiety and self-evaluation.

Our results largely confirmed our hypotheses. First, the VR interview environment successfully stimulated active social comparison among participants. Building on previous research, we incorporated "raising hands" as ambiguous social comparison information within the interview scenario. This behavior is considered a useful indicator of behavioral engagement~\cite{boheim2020student} and a naturally occurring piece of performance-related information~\cite{hasenbein2023investigating}, making it both common and plausible in interview settings. Participants consistently and quickly noticed this subtle behavior when competitors raised their hands during the quick-response phase. In the post-interview session, nearly all participants accurately recalled the number of competitors who raised their hands and their corresponding seating positions. This cue also elicited tangible responses: some participants reported increased psychological stress, faster thinking, and a compulsion to raise their hands more quickly, even when they had not yet formulated an answer.

Descriptive statistics from our eye-tracking data revealed that participants' eye movements varied with the number of competitors raising their hands. Moreover, scenarios with clearer comparison cues triggered stronger attentional responses. For instance, in the scenario where two competitors (NPCs) raised their hands, participants exhibited greater gaze duration, higher gaze frequency, and larger pupil diameter, indicating deeper information processing~\cite{mason2013eye}. This aligns with established findings that gaze direction is closely associated with the focus of attention~\cite{just1980theory} and that people attend to information that deviates from their expectations~\cite{jainta2010analyzing}. A possible explanation is that in a competitive group interview emphasizing individualism, a scenario where only one-quarter of competitors raised their hands is unusual. This finding also supports the idea that students pay more attention to peer behaviors when social comparison information clearly indicates peer achievements in extreme conditions~\cite{mussweiler2003comparison,hasenbein2023investigating}. These results indicate that participants proactively engaged in social comparison during the interview.

Second, when objective comparison information was absent, social comparison influenced participants' self-evaluation—consistent with the Big-Fish-Little-Pond Effect (BFLPE)—but it did not significantly affect their anxiety levels. The BFLPE posits that individuals of the same ability tend to have a lower self-concept in high-performing groups than in low-performing groups~\cite{marsh1987big}. In our experimental design, we manipulated the number of competitors raising their hands to control the perceived ability level of the group. NPCs who raised their hands were programmed to provide near-perfect answers, leading participants to associate the hand-raising behavior with high ability. This design encouraged participants to equate the number of hand-raisers with the group's overall ability level~\cite{hasenbein2023investigating}. Thus, we defined interview groups with two hand-raisers as "low-ability groups," where participants could perceive themselves as the "big fish." Conversely, groups with six hand-raisers were "high-ability groups," where participants were the "small fish."

Multiple regression analyses using eye-tracking indicators as dependent variables showed that, without objective comparison information, an increase in hand-raising competitors significantly increased participants’ gaze frequency and gaze duration. This suggests that a higher number of hand-raisers intensified the participants' engagement in social comparison. Furthermore, multiple regression analyses using self-evaluation as the dependent variable showed that increases in gaze duration and pupil diameter were associated with decreases in self-evaluation. In other words, the more time participants spent processing comparison information and the more cognitive resources they allocated to it, the more likely they were to lower their self-evaluations. In summary, the BFLPE was validated through the gaze duration indicator; participants in higher-performing groups spent more time observing competitors, which subsequently led to lower self-evaluations.

However, the correlation between social comparison, self-evaluation, and anxiety was not statistically significant. Regression analysis indicated that more anxious participants did not engage in social comparison more proactively, and increased social comparison did not exacerbate their anxiety. Additionally, there was insufficient evidence that anxiety was a contributing factor to reduced self-evaluation, a finding that deviates from previous research~\cite{mitchell2014experimental}.

To assess anxiety, we selected six performance-related questions from the MASI scale, a common measure for interview anxiety~\cite{mccarthy2004measuring}, and added the question, "During the interview, I was worried that my performance was worse than other candidates." While 16

Previous research highlights a similar discrepancy between reported and physiological anxiety. For example, in VR interviews, factors like question type and realism affect self-reported anxiety, while only question type and interviewer attitude affect physiological anxiety measured via skin conductance~\cite{luo2024using}. Our study did not aim to investigate external anxiety factors. To ensure ecological validity, we used authentic interview questions, maintained a high level of realism, kept the interviewer's attitude neutral, and provided participants with a question bank a week in advance. These design choices may have contributed to the relatively low anxiety levels observed and the lack of significant correlations. Nevertheless, regression analysis revealed that the covariate realism had a significant effect on anxiety; a lack of realism increased anxiety, extending Luo's findings~\cite{luo2024using}. Furthermore, the covariate intellectual self-evaluation also significantly affected anxiety, with participants who perceived themselves as less intelligent being more likely to experience interview anxiety, corroborating findings on Social Anxiety Disorder (SAD)~\cite{mitchell2014experimental}.

Additionally, objective comparison information effectively enhanced participants' engagement in social comparison, but its impact on self-evaluation remains unclear. Social comparison was originally defined as a means for individuals to evaluate themselves without objective standards~\cite{festinger1954theory}, but subsequent research shows it occurs even when such standards are present~\cite{kruglanski1990classic,klein1997objective,wood1996social}.

We used a natural "break" in the interview to create a comparative condition, observing whether the introduction of objective information (publicly disclosed test scores) influenced social comparison and self-evaluation. Descriptive statistics showed that when scores appeared above each avatar's head, participants’ gaze count, frequency, and duration on competitors increased. Regression analysis indicated that objective information amplified the positive effect of more hand-raisers on gaze frequency and duration. Notably, we observed gender differences in gaze duration, with female participants tending to observe competitors for longer. This aligns with prior research suggesting that gender can influence social comparison behavior~\cite{mitchell2014experimental} and its impact on self-evaluation~\cite{holm2020big}. However, despite this increased comparison intensity, participants' self-evaluation did not uniformly decrease as might be expected~\cite{marsh1987big}. Descriptive statistics revealed a mixed pattern: compared to the first round, participants in the two-hand-raising condition reported a significant increase in self-evaluation, those in the four-hand-raising condition reported a significant decrease, and those in the six-hand-raising condition showed almost no change.

Regression analysis showed a significant positive correlation between gaze number and self-concept, meaning participants who visually scanned more peers were more likely to give themselves a higher evaluation. Researchers suggest two possible explanations for these complex results. First, participants in the two- and four-hand-raising conditions had higher Social Comparison Orientation (SCO) scores, making their self-evaluation more susceptible to social comparison~\cite{gibbons2002drinking,buunk2005social}. Second, the public scores allowed participants to clearly identify their rank within the group, altering their judgments of their own and their competitors' abilities.

Lastly, our study did not find a significant relationship between SCO, social comparison preference, and interview anxiety. Previous studies suggest that SCO can assess preferences for upward or downward comparison~\cite{gibbons1999individual}, and individuals with higher SCO tend to prefer upward comparisons~\cite{buunk2005social}, which can lead to more negative self-evaluations~\cite{mitchell2014experimental} and anxiety~\cite{mccarthy2020exploring}. Although our regression analysis did not show significant causal relationships, descriptive statistics from the entire process showed that the four-hand-raiser condition prompted the highest SCO scores, highest anxiety, and lowest self-evaluation, which aligns with previous conclusions. Unexpectedly, this group reported the lowest tendency toward upward comparison, preferring to review "failure experiences" over "success experiences" pre-interview. Participants explained that failure experiences provided clearer lessons, aligning with findings that individuals often learn more from failure than success~\cite{lockwood2002could}.

From participants' performance and eye-tracking data, we found that when competitors raised their hands, participants focused not only on them but also on others who did not raise their hands, particularly those perceived as similar to themselves or highly regarded. This is consistent with research showing that individuals either identify with comparison targets by focusing on similarities or contrast themselves by emphasizing differences~\cite{wood1989theory}. Participants confirmed in post-interviews that they paid closer attention to these specific competitors. They also noted that their self-evaluations were not solely based on in-situ social comparisons but were also influenced by past performance and pre-interview expectations. Those with high expectations were more likely to feel frustrated and lower their self-evaluations, while those with low expectations were often more positive about their performance.

This study collected data from N=73 participants, resulting in 219 datasets across two interview rounds. Data were analyzed using descriptive statistics, ANOVA, and multiple regression in Python 3. While the data partially supported our hypotheses, we acknowledge that some regression models had low or negative adjusted R-squared values when covariates were excluded, which subsequent model corrections did not significantly improve. This was likely due to an insufficient sample size caused by the professional nature of the interview content. Future studies could lower the content threshold to recruit a larger sample. Moreover, this study relied on self-reported anxiety; future research could integrate physiological data, such as skin conductance, for a more comprehensive evaluation~\cite{luo2024using}. Lastly, to simulate a realistic group interview, we did not vary seating positions, which allowed participants to quickly familiarize themselves with the environment. However, due to interview etiquette, some participants kept their gaze fixed on the interviewer, which may have influenced some conclusions drawn from the eye-tracking data.

One point is clear: our study contributed to validating the potential of immersive VR for simulating real-world environments. To minimize VR-induced discomfort that could confound measures, we limited interview durations to 15 minutes (average 748 seconds). Post-interview, participants reported very low levels of discomfort from the equipment (M $<$ 2.0 on a 7-point Likert scale). Despite the short duration, participants in all three conditions reported relatively high levels of presence (M $>$ 5.0) and realism (M $>$ 4.6).

Furthermore, this study aims to provide practical recommendations for improving VR interview systems. Our findings suggest that user-centered systems should address job seekers’ internal needs by offering training scenarios with varied environments, competitor numbers, and ability levels. This could help desensitize anxious candidates and teach diverse skills to others. Additionally, VR’s visualization capabilities could be leveraged to provide appropriate objective comparison information, potentially mitigating the anxiety and inaccurate self-assessments caused by uncertainty.

\section{Conclusion}

This study explored the mechanisms of social comparison psychology in immersive virtual reality group interviews, providing human-centered insights into the application of VR in interview settings. The findings revealed that ambiguous social comparison information (e.g., hand-raising behavior) in VR interview scenarios significantly influenced participants’ attention to and processing of comparison information, exerting an impact on self-evaluation akin to the BFLPE. Objective comparison information (e.g., publicly disclosed written test scores) further amplified participants’ social comparison activity but mitigated the negative effects on self-evaluation by reducing uncertainty.
In our study, social comparison did not directly influence participants’ anxiety levels, as anxiety exhibited individual differences (e.g., based on intellectual self-evaluation). These findings provide theoretical support for developing more human-centered and efficient VR interview systems, with potential applications in other competitive scenarios such as debates. Our research represents an innovative step in the design of VR interview systems, particularly by addressing job seekers’ internal needs and filling gaps in current research.
As mentioned in the discussion, our study also has limitations, including a small sample size and insufficient physiological data collection. Future research could address these limitations to enable further exploration and optimization of VR interview systems.

\newpage

\bibliographystyle{abbrv-doi}

\bibliography{template}
\end{document}